\documentclass{article}

\usepackage{PRIMEarxiv}

\usepackage[utf8]{inputenc} 
\usepackage[T1]{fontenc}    
\usepackage{hyperref}       
\usepackage{url}            
\usepackage{booktabs}       
\usepackage{amsfonts}       
\usepackage{nicefrac}       
\usepackage{microtype}      
\usepackage{lipsum}
\usepackage{fancyhdr}       
\usepackage{graphicx}       
\usepackage{amsmath}
\usepackage{amssymb}
\graphicspath{{media/}}     

\newcommand{\WI}[2]{#1_{\mathrm{#2}}}
\newcommand{\Mtot}{\WI{M}{tot}}
\newcommand{\orb}{\WI{\Omega}{orb}}

\title{Non-conservative Mass Transfer \\ in the Neutron Star Stripping Model
\thanks{\textit{\underline{Citation}}: 
N.I.~Kramarev and A.V.~Yudin. Non-conservative Mass Transfer in the Neutron Star Stripping Model. \textit{Astron. Lett.} \textbf{51}, 143–150 (2025). \textbf{DOI: 10.1134/S1063773725700306.}} 
}

\author{
  N.I.~Kramarev \\
  Sternberg Astronomical Institute, Moscow State University, Moscow, 119234 Russia \\
  National Research Center ''Kurchatov Institute'', Moscow, 117218 Russia\\
  \texttt{kramarevnikita99@gmail.com} \\
   \And
   A.V.~Yudin \\
   National Research Center ''Kurchatov Institute'', Moscow, 117218 Russia\\
   \texttt{yudin@itep.ru} \\
}

\begin{document}
\maketitle

\begin{abstract}
The process of long-term stable mass transfer (or stripping) in a close neutron star binary system is possible at a sufficiently large initial asymmetry of the component masses. At the final stage of the evolution of such systems, the low-mass neutron star fills its Roche lobe, whereupon its mass is gradually transferred to the more massive component. At a certain point, the stability of the mass transfer is lost, causing
the minimum-mass neutron star to explode. In the present stripping calculations, the effect of non-conservative mass transfer has been taken into account for the first time, resulting in an increase in the duration of stable mass transfer from a few tenths of a second to a few seconds. This allows the time delay of 1.7 s between the loss of the gravitational-wave signal and the detection of the gamma-ray burst from the multimessenger event GW170817–GRB170817A to be naturally explained. The interaction of the envelope of the exploded minimum-mass neutron star with the matter ejected during non-conservative mass transfer may explain two episodes in the light curve of this gamma-ray burst.	
\end{abstract}


\section{INTRODUCTION}

The stripping model was first proposed by Clark and Eardley \cite{ClarkEardley1977} as a possible scenario for the final evolutionary stage of a neutron star (NS) binary system, which can occur when the component masses are sufficiently asymmetric. Subsequently, Blinnikov et al. \cite{Blinnikov1984,Blinnikov1990} applied this model to explain the origin of short cosmological gamma-ray bursts (GRBs).

We briefly outline the essence of the stripping mechanism (see \cite{Blinnikov2021,Blinnikov2022} for details). In an NS binary system with a large mass ratio, the two stars gradually spiral toward each other due to the emission of gravitational waves (GWs). As a result, the low-mass NS is the first to fill its Roche lobe and begins to transfer mass onto its more massive companion. The asymmetry of the system increases due to the accretion of matter, and the components recede from each other. The stripping process proceeds on a comparatively long time scale (of the order of seconds, corresponding to thousands of orbital revolutions) determined by the rate at which the system loses angular momentum through GW emission. At a certain point, the mass transfer stability is lost, whereupon the remnant of the low-mass component reaches the minimum NS mass limit (of the order of a tenth of the solar mass) on the hydrodynamic time scale and explodes to produce a rich, spherically symmetric electromagnetic transient, including the mentioned GRB.

The interest in this model increased significantly after August 17, 2017, when first the GW event GW170817 and then, with a slight delay of 1.7 s, the short GRB170817A were recorded from a pair of NSs \cite{Abbott2017,Pozanenko2018}. Shortly afterwards, the emission from the kilonova AT2017gfo associated with GW170817 was also
detected \cite{Villar2017}. The stripping model could naturally explain the main observational manifestations of this multimessenger event \cite{Blinnikov2021}: the low GRB energy, the high mass of the red component of the kilonova \cite{Siegel2019}, its
spherical symmetry \cite{Sneppen2023}, and the time delay between the GW signal loss and the GRB detection corresponding to the duration of stable mass transfer in the stripping model. The latter property of the event was reproduced by developing the
analytical approach of Clark and Eardley \cite{ClarkEardley1977} that allows the binary evolution on comparatively long time scales to be calculated. Initially, having repeated the calculations in the formulation of Clark and Eardley \cite{ClarkEardley1977} with up-to-date equations of state, we obtained a stripping time of the order of a few seconds, corresponding to the GW170817–GRB170817A delay \cite{Blinnikov2022}. However, the subsequent development of this approach by taking into account the important effect of accretion spin-up of the massive component \cite{KramarevYudin2023acc,KramarevYudin2023str} reduced significantly the duration of stable mass transfer to a few tenths of a second! This reduction occurs because the orbital angular momentum of the system is efficiently transferred to the spin angular momentum of the accretor. The uncertainties related to the initial component masses, given the weak constraints from GW observations, and the contribution from tidal spin-down of the accretor (see Figs.~3 and 4 in \cite{KramarevYudin2023str}) can increase this value only by a few tens of percent.\footnote{Note that the uncertainty of the equation of state in the range of low NS masses within the limits of agreement with terrestrial experiments can increase significantly the stripping time \cite{YudinKramarevPanov2023}.}

So far, following Clark and Eardley \cite{ClarkEardley1977}, we have used the approximation of conservative mass transfer in a binary system, where all of the energy
of the accreting matter is converted into neutrinos that leave the system unimpeded. However, the discrepancy between the calculated and observed stripping times for GW170817–GRB170817A, as well as the general motivation to refine the analytical framework, motivates us to take into account the non-conservative nature of mass transfer.

In this work, we estimate the extent to which the effect of non-conservativeness influences the evolution of an NS binary system by considering the limiting case where all of the matter being stripped from the low-mass NS surface is ejected from the
system. The structure of the paper is as follows: first we write the equations of final evolution of the NS–NS system in the most general form by taking into account the non-conservativeness, whereupon we present our calculations for the limiting
cases of completely conservative/non-conservative mass transfer. In the final section we summarize the results of our study and discuss the properties of the additional electromagnetic transient directly associated with non-conservative mass transfer, which may explain the observed double-peaked gamma-ray emission in GRB170817A and GRB190425.

\section{FORMULATION OF THE PROBLEM} 

\subsection{General Assumptions}

We formulated and justified the basic assumptions of the analytical approach being developed in \cite{KramarevYudin2023acc,KramarevYudin2023str,YudinKramarevPanov2023}. Let us list them. The NS–NS system consists of two point masses: $M_1$ (accretor) and $M_2$ (donor), where $M_1\geq M_2$. The components revolve around each other in quasi-circular orbits with the Keplerian orbital frequency
\begin{equation}
	\orb=\sqrt{\frac{G\Mtot}{a^3}},\label{O_orb}
\end{equation}
where $a$ is the distance between the NS centers of mass and $\Mtot=M_1+M_2$ is the total mass of the system. The total mass can change with time ($\WI{\dot{M}}{tot}\leq 0$) due to the possible non-conservativeness of the mass transfer during the stripping of the donor star—therein lies the main novelty of this paper compared to the previous ones.

The stars have spin angular momenta $\WI{J}{1,2}=I_{1,2}\Omega_{1,2}$, where $I_{1,2}$ are the moments of inertia and $\Omega_{1,2}$ are the spin frequencies. The NS moment of
inertia in the most general form depends not only on its mass, but also on its spin angular momentum, i.e., $I_{1,2}=I_{1,2}(M_{1,2},J_{1,2})$. The procedure for calculating
the moment of inertia (and the equatorial radius) of a rotating NS is described in detail in Appendix~B from \cite{KramarevYudin2023str}. The spin angular
momentum vectors of the stars are assumed to be perpendicular to the orbital plane. We assume that the NSs were tidally synchronized over many years of their joint evolution. Therefore, before the onset of mass transfer the stars are in corotation, i.e., $\Omega_{1,2}=\orb$. Moreover, the donor continues to be in corotation even after the onset of stripping.

\subsection{The Equation of Change in the Total Angular Momentum of the System}

The total angular momentum of the NS–NS system is the sum of the spin angular momenta of its components and the orbital angular momentum of the system:
\begin{equation}
	J_{\mathrm{orb}}=\frac{M_1 M_2}{\Mtot}a^2 \orb.\label{J_orb}
\end{equation}
The total angular momentum of the system is lost due to the emission of GWs and the ejection of matter related to the possible non-conservative mass transfer after the onset of donor stripping. The equation governing the change in the total angular momentum of the system can then be written in the most general form as
\begin{equation}
	\dot{J}_{\mathrm{orb}}+\dot{J}_1+\dot{J}_2=\dot{J}_{\mathrm{GW}}+\dot{J}_{\mathrm{ML}},\label{J_main}
\end{equation}
where the GW losses are described by the well known formula (see, e.g., \cite{LandauLifshitz1975}):
\begin{equation}
	\dot{J}_{\mathrm{GW}}=-\frac{32}{5}\frac{G}{c^5}\frac{M_{1}^2 M_{2}^2}{\Mtot^2}a^4 \orb^5.\label{J_gw}
\end{equation}
The dot here and below denotes a derivative with respect to time $t$. Since the NSs are in corotation before the onset of mass transfer, the derivatives of their spin angular momenta take the following form:
\begin{equation}
	\dot{J}_{1,2}=\Biggl[I_{1,2} \dot{\Omega}_{\mathrm{orb}}{+}\Omega_{\mathrm{orb}}\dot{M}_{1,2} \left(\frac{\partial I_{1,2}}{\partial M_{1,2}}\right)_{\!\!\! J_{1,2}}\Biggl]\beta_{1,2},\label{J12_dot}
\end{equation}
where $\beta_{1,2}{=}\left(1{-}\Omega_{\mathrm{orb}}\left(\frac{\partial I_{1,2}}{\partial J_{1,2}}\right)_{\!\! M_{1,2}}\right)^{-1}$. As we mentioned
above, Eq.~(\ref{J12_dot}) is valid for the donor NS even after the onset of stripping, when it fills its Roche lobe, while for the accretor see Eq.~(\ref{J1_acc}) below. During
stable mass transfer the average radius of the low-mass component is equal to the effective radius of the Roche lobe occupied by it: $R_2=R_{\mathrm{R}}$. We parameterize
the effective radius of the Roche lobe in accordance with \cite{Eggleton1983}:
\begin{equation}
	R_{\mathrm{R}}{=}a f(q'), \, f(q'){=}\frac{0.49 (q')^{2/3}}{0.6(q')^{2/3}{+}\ln\big[1{+}(q')^{1/3}\big]}, \label{R2_Roche}  
\end{equation}
where $q'=M_2/M_1$ is the component mass ratio. Note that below we will also use a different definition of the mass ratio, $q=M_2/\Mtot$.

After the low-mass NS fills its Roche lobe, stable mass transfer to the more massive component through the inner Lagrange point $L_1$ begins. Two accretion regimes can take place (see, e.g., \cite{LubowShu1975}). In one case, the accretion stream
hits the surface of the accretor with an equatorial radius $R_1$. In this case, the orbital angular momentum of the system $\WI{J}{orb}$ is transferred to the spin angular
momentum of the accretor $J_1$ and partially returns
back due to the tidal spin-down of the latter.\footnote{The influence of the tidal accretor spin-down effect on the evolution of an NS–NS system was investigated in our
previous paper \cite{KramarevYudin2023str}. Here, we will ignore it lest the attention of the reader be distracted.} The mass asymmetry of the system will increase with time
due to the accretion of matter, and the components will recede from each other. If the stripping process proceeds for a sufficiently long time, then the formation of an accretion disk is possible at some instant. This will occur when the point of minimum
approach of the accretion stream $\WI{R}{m}$ is greater than the equatorial radius of the accretor $R_1$. Here, we will not provide the spin-up formula in the regime of
disk accretion, since in the approximations of non-conservative/conservative mass transfer being considered below this regime does not occur. If $M_1$ at some instant becomes larger than the maximum mass of a NS with the same spin angular momentum,
then it collapses into a black hole. In this case, we will assume the accretor radius to be equal to $R_1=3 R_g$, where $R_g$ is the Schwarzschild radius of a gravitating body with a mass $M_1$. For the justification of this approach, along with the discussion of various relativistic effects, see \cite{KramarevYudin2023str,Kramarev2024}.

The accretion spin-up equation in the regime of direct impact accretion can be written in the form \cite{KramarevYudin2023acc}:
\begin{equation}
	\dot{J}_\mathrm{1} = 		
	\dot{M}_1\mathfrak{j}(q,r_1)a^2\orb, \label{J1_acc}
\end{equation}
where  $\mathfrak{j}$ is the specific angular momentum of the accreting matter in orbital units in an inertial frame dependent on the mass ratio $q=M_2/\Mtot$ and the
accretor stopping radius $r_1=R_1/a$. Using the momentum
and angular momentum conservation laws for the ''accretor–incident test particle'' system before and after the impact as well as the center-of-mass
equation, it is easy to show (see Appendix~1) that the angular momentum in an inertial frame is
\begin{equation}
	\mathfrak{j}(q,r_1)=j(q,r_1) + r_1^2,\label{j_acc}
\end{equation}
where $j$ is the specific angular momentum of the accreting matter in orbital units in a rotating frame, for which we previously \cite{KramarevYudin2023acc} selected an approximation to better than 1\%. 

\subsection{Allowance for the Non-conservativeness}

Let us describe the main novelty of this paper — allowance for the loss of angular momentum due to the possible non-conservativeness during the mass
transfer attributable to supercritical accretion. As shown in Appendix~2, the optical depth of the medium of matter falling onto the NS is much greater than unity. Therefore, the pressure of the high-energy photons generated during accretion can eject part of
the matter from the system. We assume that this mass loss is spherically (or at least cylindrically) symmetric. The corresponding angular momentum loss can then be estimated as
\begin{equation}
	\dot{J}_{\mathrm{ML}}=(1-\chi)\dot{M}_2 a_1^2 \orb,\label{J_ml}
\end{equation}
where $a_1=a q$ is the distance from the system’s center of mass to the accretor center and $0\leq\chi{=}{-}\dot{M}_1/\dot{M}_2 \leq 1$ is the fraction of the matter falling onto the accretor ($\chi=1$ for conservative mass transfer). The remaining fraction of the accreting matter, $1-\chi$, is swept out from the system by radiation. Following Clark and Eardley \cite{ClarkEardley1977}, in all our previous papers we assumed that the overwhelming fraction of the accreting matter energy would be radiated in the form of neutrinos leaving the system unimpeded. Most of the energy can also go into heating the accretor
itself. Without going into the physical details of the impact of accreting matter on the NS surface and the neutrino and photon emission mechanism, let us estimate the influence of non-conservativeness in comparison with completely conservative mass transfer. For this purpose, consider the limiting case of non-conservativeness where all of the accreting matter is swept out from the system. Strictly speaking, it is impossible for \emph{all} of the accreted matter to be ejected: at least a small fraction ($\chi \ll 1$) must reach the surface of the massive NS to generate the intense radiation that makes the mass transfer non-conservative. However, in our phenomenological consideration, while estimating the extent of the effect of non-conservativeness, we will assume in one of the limiting cases that $\chi=0$. Thus, specifying the parameter $\chi=1$ or $\chi = 0$, we close the system of evolution equations for the NS–NS system during stripping.

\subsection{The Mass Transfer Stability Criterion}

For stable mass transfer the size of the Roche lobe (\ref{R2_Roche}) of the low-mass component must increase more rapidly than its radius, $\Delta R_{\mathrm{R}} \geq \Delta R_2$ (see, e.g., \cite{Blinnikov2022}). Using Eqs.~(\ref{O_orb})--(\ref{J_ml}) and defining $\chi=-\dot{M}_1/\dot{M}_2$, we can obtain the mass transfer
stability criterion as a function of the system’s parameters. Since the corresponding general formula is cumbersome, in this paper we will consider its limiting
cases. For $\chi=1$ this criterion can be written as
\begin{equation}
	\frac{d \ln R_2}{d \ln M_2}\geqslant \frac{d \ln f}{d \ln q } 
	- 2\frac{1{-}2q{-}\mathfrak{j}{+}\frac{\beta_2}{ a^2}\left(\frac{\partial I_2}{\partial M_2}\right)_{\!\! J_2}}{1{-}q{-}\frac{3\beta_2 I_2}{ M_2 a^2}}. \label{stabConservative}
\end{equation}
This expression defines the boundary between the merging and stripping scenarios and the time of stable mass transfer in the stripping mechanism under
the assumption of conservative mass transfer. We first derived Eq.~(\ref{stabConservative}) in \cite{KramarevYudin2023str}.

For completely non-conservative mass transfer ($\chi=0$) the mass transfer stability criterion can be written as
\begin{equation}
	\frac{d \ln R_2}{d \ln M_2}\geqslant \frac{d \ln f}{d \ln q }\left(1{-}q\right) 
	- \frac{2{-}3q{-}q^2{+}\frac{2\beta_2}{ a^2}\left(\frac{\partial I_2}{\partial M_2}\right)_{\!\!J_2}\! \! \! {+}\frac{q\beta_2 I_2}{ M_2 a^2}}{1{-}q{-}\frac{3\beta_2 I_2}{ M_2 a^2}}. \label{stabNonConservative}
\end{equation}
Since in this case all of the accreting matter is ejected from the system, the spin-up of the massive component has no effect on the evolution of the system, i.e., $\dot{J}_\mathrm{1} = 0$.

\section{RESULTS OF OUR CALCULATIONS}

\begin{figure}[h]
	\centering
	\begin{minipage}[h]{0.49\linewidth}
		\includegraphics[width=\columnwidth]{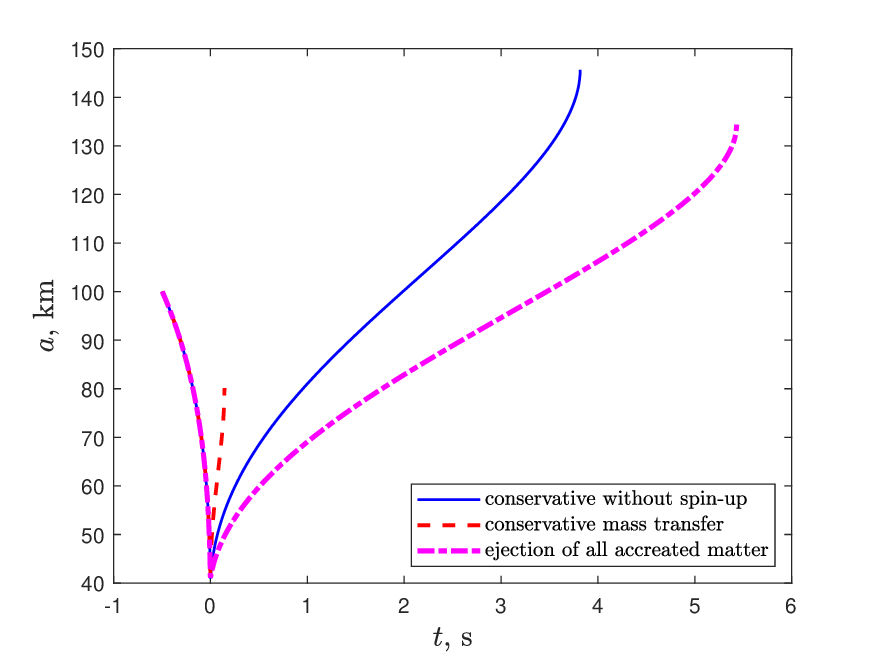}
		\caption{Evolution of the distance between the components with masses $M_1=1.7\WI{M}{\odot}$ and $M_2=0.8\WI{M}{\odot}$ for different mass
		transfer regimes. The blue solid line corresponds to the case of conservative mass transfer without accretion spin-up. The red dashed line illustrates conservative mass transfer (with spin-up). The purple dash–dotted line indicates the evolution of the distance under the assumption that all of the accreting matter is ejected from the system. For details, see the text.}
		\label{Fig1distance}
	\end{minipage}
	\hfill
	\begin{minipage}[h]{0.49\linewidth}
		\includegraphics[width=\columnwidth]{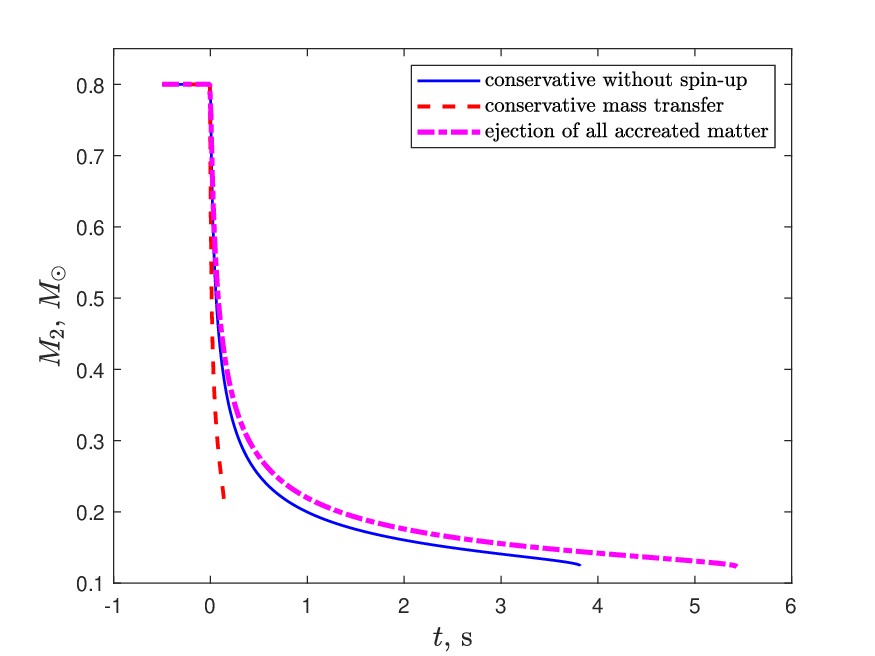}
		\caption{Evolution of the donor mass during stable mass transfer. The designations and initial conditions are the same as those for Fig.~\ref{Fig1distance}.}
		\label{Fig2mass}
	\end{minipage}
\end{figure}

Let us consider how allowance for the non-conservativeness influences the stripping process in an NS–NS binary system. Figures~\ref{Fig1distance} and \ref{Fig2mass} show
the changes in the distance and the donor mass with time. The red dashed curve represents completely conservative mass transfer ($\chi=1$), while the purple
dash–dotted line corresponds to the situation where all of the matter falling onto the accretor is ejected from the system ($\chi=0$). For comparison, we also
present the case of conservative mass transfer (blue curve), but without accretion spin-up, i.e., $\dot{J}_\mathrm{1} = 0$ after the onset of stripping. In all our calculations we use the same BSk22 equation of state \cite{Pearson2018}. Two NSs with initial masses $M_1=1.7\WI{M}{\odot}$ and $M_2=0.8\WI{M}{\odot}$ approach each other to
a distance $a\approx40\,$km, where the low-mass NS fills its Roche lobe. Before this instant of time ($t=0$) all three curves coincide, as it must be. However, during
stable mass transfer the curves diverge significantly. For conservative mass transfer (red curve) we see that the components move apart to a distance $a\approx80\,$km in a few tenths of a second, whereupon the stability condition (\ref{stabConservative}) is violated. For the other two cases the stripping process lasts for seconds. As can be seen
from Fig.~\ref{Fig2mass}, at the instant when the mass transfer stability is lost, the donor masses differ greatly: for the red curve $M_2\approx0.2\WI{M}{\odot}$, while for the blue and purple ones it is close to the minimum NS mass limit.

Note separately that the time of stable mass transfer for the non-conservative case is longer than that for the conservative one, where the accretion spin-up
is ignored. Although in both cases $\dot{J}_\mathrm{1}=0$, their evolutions slightly differ. In particular, since $\Mtot=\mathrm{const}$ for the blue curve and $\WI{\dot{M}}{tot}< 0$ for the purple one, the mass ratio $q=M_2/\Mtot$ that enters into
the stability criteria (\ref{stabConservative}) and (\ref{stabNonConservative}) changes differently. This also affects the fate of the accretor. For the initial masses of the components being used we obtain a massive NS at the end in the non-conservative case
and a black hole surrounded by an accretion disk in the conservative one (without spin-up), which can give rise to a jet and, accordingly, an extended GRB emission (see, e.g., \cite{Barkov2011}). Interestingly, in the case of conservative mass transfer,
but already with accretion spin-up, the massive NS will not collapse into a black hole, because the maximum limiting mass for rotating NSs increases
(see Appendix~B in \cite{KramarevYudin2023str}).

\section{DISCUSSION AND CONCLUSIONS}

We showed that non-conservativeness could lead to a significant increase in the duration of stable mass transfer from a few tenths of a second to a few seconds. Recall that observationally the stripping time corresponds to the time interval between the
GW signal loss and the GRB detection (1.7 s for the GW170817–GRB170817A event). Non-conservative mass transfer also affects the ultimate fate of the accretor, as well as the mass of the accretion disk and the ejected matter. It can also lead to an additional electromagnetic signal, along with the transient during the explosion of a minimum-mass NS discussed in our previous papers.

Let us discuss the latter in more detail following the observational data on the structure of GRB170817A presented in \cite{Pozanenko2018}. The pulse of this GRB consisted of two peaks: a ''hard'' peak of gamma-ray photons that emerged
1.7 s after the GW signal loss and a ''softer'' one that appeared in another $\sim\!\!2$ s. The isotropic radiated energy $\WI{E}{iso}$ was $\WI{E}{iso}\approx 5\cdot 10^{46}$~erg in the first peak and $\WI{E}{iso}\approx 1.5\cdot 10^{46}$~erg in the second one. Within the stripping model, the sequence of events may be as follows: non-conservative mass transfer begins at the instant of closest approach of the NSs corresponding to the GW emission peak. In this case, the \emph{bulk} of the mass is transferred (and ejected from the system) within the first tenths of a
second (see, e.g., \cite{ClarkEardley1977, KramarevYudin2023str}), while in the remaining time the accretion rate is relatively low. Within 1.7 s the low-mass
NS reaches the lower mass limit and explodes to produce the first gamma-ray emission pulse. Its matter with a speed $\sim\!\!10$\% of the speed of light \cite{Blinnikov1990, Yudin2022} catches up with the matter ejected before this time approximately in 2 s. This allows us to estimate the ejection velocity, $\WI{v}{ej}\approx 0.05c$, — this important parameter should be independently calculated in the future
physical model of non-conservative mass transfer within the stripping scenario. The interaction of these ejecta produces the second, considerably softer burst of radiation. The underlying physics here is very similar to the interaction of the shock from a supernova with its ejecta (shock interacting SN) — a process invoked to explain the so-called superluminous supernovae \cite{Sorokina2016}.

It is also interesting to note that GRB190425, corresponding
to the GW signal S190425z from the second known pair of merging NSs \cite{Abbot2020}, also shows many features similar to the GRB170817A event \cite{Pozanenko2020}. In particular, two peaks observed approximately 0.5 and 5.9~s after
the GW emission maximum are distinguishable in its structure. These values agree well with the scenario described above. On the other hand, the estimated range of isotropic energies for this event, $\WI{E}{iso}\approx (2.2\cdot 10^{47}-6.7\cdot 10^{48})$~erg, is approximately an order of magnitude higher than the GRB energetics in the stripping model \cite{Blinnikov1990}. Here, it is pertinent to recall that the stripping scenario
by no means purports to explain the \emph{entire} population of GRBs. However, it can also be noted that the interaction of the matter ejected during the low-mass
NS explosion (its kinetic energy is $10^{51}$~erg, see \cite{Blinnikov1990}) with the surrounding matter may well increase significantly the old estimates of the
GRB energetics in the stripping model.

Recall that if the initial total mass of the system is greater than the limiting maximum NS mass, then the massive NS can collapse into a black hole during stripping \cite{KramarevYudin2023str}. Moreover, at the final mass transfer stages, when the binary
components move apart to a fairly large distance ($\sim\!\!100$~km), the regime of disk accretion onto the black hole can be established, giving rise to a classical
jet. The additional jet radiation, along with the mechanism of the interaction between two quasi-spherical ejecta described above, makes the palette of observational manifestations of the stripping model even more varied than we have thought previously.

To summarize, it can be said that the non-conservativeness of the mass transfer in the stripping model is an extremely important effect and can affect the main parameters of the model. However, the construction of a physical model of non-conservative mass transfer that must answer the questions about the fraction of the ejected matter, its velocity, etc. is required to make a proper quantitative allowance for this
effect. The question of the energy balance of the accreting matter is of primary importance: what fraction of its released gravitational energy goes into neutrino radiation, photons, the heating of the accretor, the kinetic energy of the ejected matter, etc. In the completely conservative case, following \cite{ClarkEardley1977}, it was thought that all of the energy goes into neutrinos. In principle, it is not surprising if most of the energy will indeed go via this channel, as the neutrino luminosity is known to increase with the accretion rate onto the NS (see, e.g., \cite{Mushtukov2018}). However, even a small remaining fraction in the energy balance can be more than enough for the mass transfer to be strongly non-conservative. We are currently working on solving this problem.

\section*{\textit{APPENDIX~1.} THE ANGULAR MOMENTUM OF THE ACCRETING MATTER \\ IN DIFFERENT FRAMES}

Let us obtain the relation between the angular momentum of an accreting particle in the inertial and synchronously rotating frames, i.e., let us derive Eq.~(\ref{j_acc}). Let the velocity of an incident particle with a mass $\triangle M_1$ in the rotating frame when hitting the surface of the massive component be given by the vector $\vec{V}$, while the velocity of the accretor itself before the impact be zero. In the inertial frame the velocities $[\vec{\Omega}_{\mathrm{orb}}\times \vec{D}]$ and $[\vec{\Omega}_{\mathrm{orb}}\times \vec{a}_1]$, respectively, where the
vectors $\vec{D}$ and $\vec{a}_1$ are drawn from the system’s center
of mass to the particle impact point and the accretor’s center of mass, are added to them. Below, we will also need the particle radius vector relative to the
accretor’s center: $\vec{R}_1=\vec{D}-\vec{a}_1$.
Let us write the momentum conservation law for the particle and the accretor before and after the impact:
\begin{equation}
	M_1[\vec{\Omega}_{\mathrm{orb}}\times \vec{a}_1]+\triangle M_1\left(\vec{V}+[\vec{\Omega}_{\mathrm{orb}}\times \vec{D}]\right)
	=(M_1{+}\triangle M_1)\vec{V}^{\mathrm{f}}_1,\label{impuls}
\end{equation}
where $\vec{V}^{\mathrm{f}}_1$ is the velocity of the star after the impact. In
turn, the angular momentum conservation law gives
\begin{equation}
	M_1\big[\vec{a}_1{\times}[\vec{\Omega}_{\mathrm{orb}}{\times} \vec{a}_1]\big]{+}\triangle M_1\big[\vec{D}{\times}(\vec{V}{+}[\vec{\Omega}_{\mathrm{orb}}{\times} \vec{D}])\big]
	=(M_1{+}\triangle M_1)[\vec{a}^{\mathrm{f}}_1{\times} \vec{V}^{\mathrm{f}}_1]{+}\triangle\vec{J}_1,\label{ugl_moment}
\end{equation}
where $\vec{a}^{\mathrm{f}}_1$ is the position of the accretor with the mass $M_1{+}\triangle M_1$ after the impact and $\triangle\vec{J}_1$ is the change in
its spin angular momentum.

In our previous paper \cite{KramarevYudin2023acc}, when deriving the formula for the relation between the angular momenta of the accreting particle in the inertial and rotating frames, we ignored the change in the accretor’s center of mass, i.e., assumed
that $\vec{a}^{\mathrm{f}}_1=\vec{a}_1$. From the condition for the center of
mass of the ''accretor–incident test particle'' to be conserved, the final position of the accretor can be written as
\begin{equation}
	\vec{a}^{\mathrm{f}}_1=\vec{a}_1+\frac{\triangle M_1}{M_1}\vec{R}_1. \label{center_mass}
\end{equation}
Using the momentum (\ref{impuls}) and angular momentum (\ref{ugl_moment}) conservation laws and taking into account the change in the accretor’s position after the impact (\ref{center_mass}), we obtain the change in the spin angular momentum of the accretor:
\begin{equation}
	\triangle\vec{J}_1=\triangle M_1[\vec{R}_1\times \vec{V}]+\triangle M_1 \vec{\Omega}_{\mathrm{orb}}R_1^2. \label{spinUp}
\end{equation}
As in our previous derivation, the first term here is the angular momentum of the particle with a mass $\triangle M_1$ in the rotating frame associated with the accretor.
The second term is its kinematic correction whose form was simplified, since it no longer depends on the position of the particle impact on the accretor’s surface. The spin angular momentum of the accretor is parallel to the orbital angular momentum of the
system. By comparing the derived formula (\ref{spinUp}) with the spin-up equation in the case of direct impact accretion (\ref{J1_acc}), we obtain the sought-for formula for the
relation between the angular momenta of the accreting matter in different frames (\ref{j_acc}).

\section*{\textit{APPENDIX~2.} THE MEAN FREE PATH AND THE DIFFUSION TIME \\ OF THE PHOTONS EMITTED DURING ACCRETION}

Let us estimate the mean free path of the high-energy photons being generated through the direct impact of accreting matter on the NS and their diffusion time in two possible approximations. In the first case, the matter falling onto the NS remains inside
its Roche lobe with a characteristic radius $\WI{R}{R}$. The average density of the medium inside the Roche lobe is then $\rho=3 \WI{M}{acc} / (4 \pi \WI{R}{R}^3)$, where $\WI{M}{acc}$ is the mass of the accreted matter. In the other case, the matter
is ejected from the system and forms an expanding quasi-spherical shell of radius $R=\WI{v}{ej}\WI{t}{str}$, where $\WI{v}{ej}$ is the shell expansion velocity and $\WI{t}{str}$ is the time from the onset of stripping. The average density of the
medium inside the shell is $\rho=3 \WI{M}{acc} / (4 \pi \WI{v}{ej}^3 \WI{t}{str}^3)$. The
mean free path of photons with an energy $\WI{E}{\gamma}$ in a
medium with a density $\rho$ and an electron fraction $\WI{Y}{e}$ can be written as
\begin{equation}
	\WI{l}{\gamma}=\frac{\WI{m}{u}}{\WI{Y}{e} \, \rho \, \WI{\sigma}{KN}(\varepsilon)}, \label{l_gamma}
\end{equation}
where $\WI{\sigma}{KN}(\varepsilon{=}\WI{E}{\gamma}/\WI{m}{e} c^2)$ is the total cross section for the scattering of photons by an electron with mass $\WI{m}{e}$ defined by the well-known Klein–Nishina formula (see, e.g., \cite{BerestetskiiLifshitz1971}). Using the total Klein–Nishina cross section $\WI{\sigma}{KN}(\varepsilon)$ instead
of the Thomson one $\WI{\sigma}{T}$ is justified by the high virial temperature of the matter \cite{ClarkEardley1977} and, accordingly, by the high energy of the photons
being emitted: $\WI{E}{\gamma}{\sim} \WI{k}{B} T{\sim} G \WI{M}{NS}  \WI{m}{u}/\WI{R}{NS} {\sim} 10 \, \text{MeV}$. For definiteness, we will use $\WI{\sigma}{KN}(\varepsilon{=}20)\approx 0.075 \, \WI{\sigma}{T}$. For the cases where the accreting matter is retained inside the Roche lobe (below designated in
the formulas by the subscript <<$\mathrm{R}$>>) and is ejected from the system (the subscript <<$\mathrm{ej}$>>) under consideration we obtain
\begin{subequations}
	\begin{align}
		\WI{l}{\gamma, \, R}\!\!&= \!6.97 \! \cdot \!  10^{-12} \text{cm}\, \frac{0.1}{\WI{Y}{e}} \! \left(\frac{\WI{R}{R}}{10\, \text{km}}\right)^{3}\!  \frac{0.1 \WI{M}{\odot}}{\WI{M}{acc}},\\
		\WI{l}{\gamma, \, ej}\!\!&=\!1.88 \! \cdot \! 10^{-4} \text{cm}\, \frac{0.1}{\WI{Y}{e}} \! \left(\frac{\WI{v}{ej}}{0.1 c}\right)^{3} \!\! \left(\frac{\WI{t}{str}}{0.1 \, \text{s}}\right)^{3} \! \frac{0.1 \WI{M}{\odot}}{\WI{M}{acc}}.
	\end{align}
\end{subequations}
Clearly, the optical depth of the medium $\tau{\sim} R / \WI{l}{\gamma}$ in
both cases is much greater than unity. Therefore, it is interesting to also estimate the diffusion time of the leakage of photons through the medium $\WI{t}{\gamma}=3 R^2 / (\WI{l}{\gamma} c)$ in the approximations under consideration:
\begin{subequations}
	\begin{align}
		\WI{t}{\gamma, \, R}\!\!&= \! 4.55 \! \cdot \! 10^{5} \, \text{yr}  \left(\frac{\WI{Y}{e}}{0.1}\right) \frac{ \WI{M}{acc}}{0.1\WI{M}{\odot}}\left(\frac{10\, \text{km}}{\WI{R}{R}}\right),\\
		\WI{t}{\gamma, \, ej}\!\!&= \! 1.52 \! \cdot \! 10^{3} \, \text{yr}  \left(\frac{\WI{Y}{e}}{0.1}\right) \frac{ \WI{M}{acc}}{0.1\WI{M}{\odot}}\left(\frac{0.1 c}{\WI{v}{ej}}\right)\frac{0.1\, \text{s}}{\WI{t}{str}}.
	\end{align}
\end{subequations}
In both cases, the diffusion time of the leakage of photons through the ejected matter exceeds the characteristic time of stable mass transfer by many orders of magnitude, $\WI{t}{\gamma} {\gg} \WI{t}{str}$.

\section*{ACKNOWLEDGEMENTS}

We are grateful to M.V.~Barkov and A.S.~Pozanenko who drew our attention to the possible observational manifestations of the stripping model being developed related directly to the accretion of matter onto the massive component.

\section*{FUNDING}

The work of N.I.~Kramarev was supported by the ''BASIS'' Foundation for the Development of Theoretical Physics and Mathematics (project no.~22-2-10-11-1).

\bibliographystyle{unsrt}  
\bibliography{references}

\end{document}